\documentclass[prd, aps, superscriptaddress, preprintnumbers, twocolumn, floatfix, nofootinbib, fleqn, citeautoscript]{revtex4-2}
\pdfoutput=1

\usepackage{amsfonts}
\usepackage{amsmath}
\usepackage{amssymb}
\usepackage{bm}
\usepackage{dcolumn}
\usepackage{graphicx}
\usepackage[latin1]{inputenc}
\usepackage{latexsym}
\usepackage{rotating}
\usepackage{hyperref}
\hypersetup{colorlinks, urlcolor=blue, citecolor=blue}
\usepackage{graphicx}
\usepackage{xcolor}
\usepackage{physics, bm} % Added by Vincent

\newcommand\be{\begin{equation}}
\newcommand\bea{\begin{eqnarray}}
\newcommand\ee{\end{equation}}
\newcommand\eea{\end{eqnarray}}

\renewcommand{\[}{\left[}

\newcommand{\A}{\mu}
\newcommand{\B}{\nu}
\newcommand{\Hu}{\mathcal{H}}

\def\doi{http://doi.org}

\begin{document}

\title{Back-Reaction of Long-Wavelength Cosmological Fluctuations as Measured by a Clock Field}

\author{Vincent Comeau}
\email{vincent.comeau@mail.mcgill.ca}
\affiliation{Department of Physics, McGill University, Montr\'{e}al, QC, H3A 2T8, Canada}

\author{Robert Brandenberger}
\email{rhb@physics.mcgill.ca}
\affiliation{Department of Physics, McGill University, Montr\'{e}al, QC, H3A 2T8, Canada}
\affiliation{Institute for Theoretical Physics, ETH Z\"{u}rich, Switzerland}

\date{\today}

%%%%%%%%%%%%%%%%%%%%%%%%%%%%%%%%%%%%%%%%%%%%%%%%%%%%%%%%%%%%%%%%%%%%%%%%%%%%%%%%%%%%%%%%%%%%%%
\begin{abstract} 
 
 We consider the back-reaction of cosmological fluctuations on the local expansion rate averaged over a space-like hypersurface of constant value of a clock field. We show that in the infrared limit, the fluctuations lead to a decrease in the average expansion rate, measured at a fixed value of the clock field, compared to what would be obtained in a homogeneous universe. We work in the context of Einstein gravity coupled to perfect fluid matter.
 
\end{abstract}
%%%%%%%%%%%%%%%%%%%%%%%%%%%%%%%%%%%%%%%%%%%%%%%%%%%%%%%%%%%%%%%%%%%%%%%%%%%%%%%%%%%%%%%%%%%%%%

\maketitle

\section{Introduction}
 
In cosmologies with an accelerated phase of expansion, cosmological fluctuation modes continuously exit the Hubble horizon. If this phase lasts long, as it does in standard inflationary universe models, then over time a large phase space of super-Hubble modes builds up. It is an important question to ask whether these modes can have a locally measurable gravitational effect.  In the same way that the gravitational effect of matter which falls into a black hole remains visible to an observer located outside the black hole horizon, we can ask if the gravitational effect of matter fluctuation modes with wavelengths larger than the Hubble horizon can be detected by an observer located inside of the Hubble patch. This is the question of {\it back-reaction} of long wavelength cosmological perturbations.

The back-reaction of super-Hubble gravitational waves was first studied by Tsamis and Woodard \cite{WT} who found that the back-reaction effect tends to decrease the expansion rate. In the presence of matter, the cosmological fluctuations (fluctuations of matter coupled to the induced metric perturbations) generically have a larger effect than that of the gravitational waves.  In \cite{ABM}, a formalism was suggested to study the back-reaction of super-Hubble cosmological fluctuations on the background metric.  This back-reaction effect was described by an effective energy-momentum tensor by which the perturbations affect the background metric\footnote{This energy-momentum tensor is second order in the amplitude of the cosmological fluctuations, but it is phase-space-enhanced in that an increasing phase space of super-Hubble modes contributes.}. It was found that at a fixed background time $t$, the effect of the fluctuations leads to a decrease in the expansion rate
\be
\expval{H}_t \,\, < \,\, H_0(t) \, , 
\ee
where $\expval{H}_t$ is the expansion rate in the presence of fluctuations averaged over the constant $t$ surface, as indicated by the subscript, and $H_0(t)$ is the background expansion rate. In fact, the equation of state of this back-reaction effect corresponds to that of a negative contribution to the cosmological constant, which leads to the conjecture \cite{RHBbrRev} that the back-reaction of cosmological fluctuations might lead to a dynamical relaxation of the cosmological constant\footnote{See also \cite{Mottola} for other approaches towards the dynamical relaxation of the cosmological constant via fluctuations.}.  

However, Unruh \cite{Unruh} asked the key question as to whether the effect discussed in \cite{ABM} could be locally measurable, or whether it could be undone by a second-order time rescaling. It was then shown \cite{GG1}\footnote{See also \cite{AB, AW, Senatore, us} for works reaching the same conclusion.} that, in the absence of entropy fluctuations, the back-reaction effect of super-Hubble fluctuations modes is indeed a gauge artefact. More specifically,  in the case when matter is given by a single scalar field $\varphi$, then 
\be
\expval{H}_{\varphi} \, = \, H_0(\varphi) \, ,
\ee
where the left-hand side is the expansion rate averaged over the constant $\varphi$ surface.

In the real world, however, there are many components to the matter. For example, in Standard Big Bang cosmology there is both cold (i.e. pressureless) matter and radiation (the Cosmic Microwave Background or CMB). At late times, the cold matter dominates the energy budget, but time is measured in terms of the temperature of the CMB, and there are fluctuations of the distribution of the cold matter from the point of view of the constant CMB temperature surfaces. Thus, the question we should ask is whether there is a back-reaction effect of long wavelength fluctuations from the point of view of the constant temperature surfaces, i.e. we should be asking whether the average of the local expansion rate at a fixed temperature $T$ in the presence of cosmological fluctuations is the same or different than the background expansion rate at the same temperature.

Similarly, in the early universe there are many matter fields. In addition to the field $\varphi$ which dominates the energy density of the background (e.g. the ``inflaton'' field during a period of cosmological inflation \cite{Guth}), there are the usual matter fields. We will take one of these to be the {\it clock field} $\chi$. In the following we will show that, for reasonable values of the equation of state of the background,  the presence of super-Hubble fluctuations leads to a back-reaction effect which decreases the value of the spatially-averaged expansion rate,
\be
\expval{H}_\chi \,\, < \,\, H_0(\chi) \, ,
\ee
where the subscript $\chi$ indicates that the average is taken over surfaces where the clock field $\chi$ is constant. Thus, the back-reaction of super-Hubble fluctuations is real.

The fact that the back-reaction of long wavelength fluctuations has a non-vanishing effect in models with more than one scalar field was first shown in \cite{GG2}.  A more detailed analysis \cite{Vacca} showed that the back-reaction leads to a decrease in the locally measured expansion rate and hence counteracts any pre-existing cosmological constant. This result was confirmed in \cite{Leila}. While the analyses of \cite{GG2, Vacca} are to leading order in perturbation theory,  in \cite{Leila} the analysis was done to all orders in perturbation theory, but to leading order in the gradient expansion of \cite{AB}.

The past analyses of the back-reaction of super-Hubble fluctuation modes were usually done in the context of scalar field matter.  In this note we revisit the issue, taking the dominant matter to be a perfect fluid.  We will consider only the terms which are dominant in the infrared limit.

\section{Formalism}

We will work to leading order in the cosmological perturbations\footnote{See \cite{MFB} for a review of the theory of cosmological perturbations, and \cite{RHBfluctsRev} for a shorter discussion.}. We will consider general metric perturbations $h_{\A\B}$, although it will become apparent that only the scalar modes play a role in our calculation. Moreover, as we will explain later, it makes sense to express the perturbations in the uniform-density gauge, in which the perturbation to the total energy density is taken to be zero, $\delta \rho = 0$.

Since the contribution of the background spatial curvature in the present universe is measured to be negligible, we will consider a spatially flat background metric. The space-time metric therefore takes the form
\be
g_{\A\B} \, = \, a^2(\eta) \bigl( \eta_{\A\B} + h_{\A\B}(\eta, \bm{x}) \bigr) \, 
\ee
where $\eta$ is the background conformal time, $\bm{x}$ are the comoving spatial coordinates, $a(\eta)$ is the cosmological scale factor, and $\eta_{\A\B}$ is the Minkowski metric with the signature $\eta_{00} = -1$.  The background expansion rate in comoving coordinates is denoted by ${\cal{H}} = \frac{a'}{a}$. In the following, a prime will denote the derivative with respect to conformal time, and quantities with subscript $0$ are background quantities. 

We will model the background matter as a multi-component perfect fluid with total energy density $\rho$, and pressure $p$. The equation of state parameter $w$ of the background is defined as
\be
w \, = \, \frac{p}{\rho} \, .
\ee
For cold matter we have $w = 0$, for radiation $w = \frac13$, and for a background which yields inflationary expansion $w \simeq -1$. 

More specifically, we consider a perfect fluid with at least two components, one whose energy density dominates in the background, and another whose energy density, while being small, can serve as a clock field $\chi$. The clock field $\chi(\bm{x}, \eta)$ is taken to be a monotonic function of the background energy density of one of the fluid components (not the dominant one), in the same way that the CMB temperature $T(\bm{x}, \eta)$ is a monotonic function of the background energy density of photons. If $\rho_\chi$ denotes the energy density of the clock field in the perturbed space-time, and $\rho_{0\chi}$ its background value, our definition for the clock field can be stated as
\be \label{clock-field}
\rho_\chi(\bm{x},\eta) \, = \,\rho_{0\chi}\big(\chi(\bm{x}, \eta)\big).
\ee

The dominant fluid component has a normalized four-velocity $u^\A$, with $u^\A u_\A = -1$. In general, we would also need to specify the four-velocity of the clock field. However, the difference between these two velocities does not contribute to our calculation, in part because of the long-wavelength limit considered later in this paper. 

The local expansion rate $H(\bm{x}, \eta)$ is given by the divergence of the four-velocity of the dominant fluid component,
\be
H({\bm{x}}, \eta) \, = \, \frac{1}{3} \nabla_\A u^{\A} \, = \, 
 \frac{1}{3\,\sqrt{-g}} \,\partial_{\A} \big(\sqrt{-g} \,u^{\A} \big) \, ,
\ee
where $g$ is the determinant of the metric. In the following, our goal will be to compute the average value of the local expansion rate over a surface of constant clock field $\chi$. The average of a scalar function $F$ over such a surface is given by
\be
 \expval{F}_\chi (\chi_0) \, = \, \frac{\int \dd[3]{x} \sqrt{\gamma_\chi} \,F}{\int \dd[3]{x} \sqrt{\gamma_\chi}} \, ,
\ee
where the integrals are over the surface of constant clock field $\chi(\bm{x}, \eta) = \chi_0$, and $\gamma_\chi$ is the determinant of the induced metric on this surface. We wish to compare this average with the corresponding average over a surface where the total energy density of the fluid is held constant, namely,
\be
\expval{F}(\chi_0) \, = \, \frac{\int \dd[3]{x} \sqrt{\gamma} \,F}{\int \dd[3]{x} \sqrt{\gamma}} \, ,
\ee
where $\gamma$ is the determinant of the metric of such a surface. Since we have chosen to write the scalar metric perturbations in the uniform-density gauge, in which $\delta \rho = 0$, the surfaces of constant energy density and background conformal time $\eta$ coincide with each other. Hence, the induced metric on these surfaces is, at the first order,
\be
\gamma_{ij} \,=\, a^2(\eta)\big( \delta_{ij} + h_{ij}\big)\,,
\ee
and so,
\be\label{ind1}
\sqrt{\gamma} \,=\, a^3(\eta)\left( 1 + \frac12 h_{ii}\right)\,.
\ee

However, these surfaces differ from the ones where the clock field $\chi$ is constant. It is useful to define a new conformal time $\eta_\chi=\eta_\chi\big(\chi(\bm{x},\eta)\big)$, which is constant when the clock field is constant, and differs from the background conformal time by a small amount,
\be
 \delta \eta({\bm x}, \eta) \, = \,  \eta - \eta_\chi({\bm x}, \eta)  \,.
\ee
In a companion paper \cite{Comeau}, we show that the conformal time perturbation $\delta \eta$ is directly related to the entropy fluctuations of the fluid. In particular, it vanishes for purely adiabatic perturbations. In terms of this quantity, the induced metric on the surfaces of constant clock field has a determinant
\begin{align}\begin{split}\label{ind2}
\sqrt{\gamma_\chi} \,&=\, a^3(\eta)\left( 1 + \frac12 h_{ii}\right) \\
&=\, a^3(\eta_\chi) \left(1 + 3\Hu\,\delta\eta + \frac12 h_{ii}\right)\,.
\end{split}\end{align}

A detailed analysis presented in the same companion paper \cite{Comeau} shows that, at the second order in the perturbations, the average values of a scalar function $F$ are related to one another by
\begin{align}\begin{split}\label{main}
\hspace{-6.5pt} \frac{\expval{F}_{\chi} - \expval{F}}{F_0} \,&=\, \frac{F'_0}{F_0} \expval{\delta\eta}_{\chi} + \left(\frac{F'_0}{F_0} + 3\Hu\right)\expval{\delta F \delta\eta} \\[7pt]
&\hspace{10pt} + \langle\delta F'\,\delta\eta\rangle 
 - 3\Hu \expval{\delta F}\expval{\delta\eta} \\[6pt]
&\hspace{10pt} + \frac{F''_0}{2F_0}\,\langle(\delta\eta)^2\rangle 
 \,,
\end{split}\end{align}
where $F_0$ is the background value of the function, and $\delta F$ its perturbation,
\be
  F(\bm{x}, \eta) \, = \, F_0(\eta)\big( 1 +  \delta F (\bm{x}, \eta) \big) \,.
 \ee
To obtain (\ref{main}), we use the expressions (\ref{ind1}) and (\ref{ind2}) for the determinants of the induced metrics, and expand the average values of $F$ up to the second order in the perturbations. No other assumption is necessary for the derivation of this result. In particular, we did not take the long-wavelength limit of the metric perturbations, nor did we assume that the energy density of the clock field is small compared to the other fluid components.

\section{Analysis} 

We can now apply the main formula (\ref{main}) to obtain the average value of the local Hubble expansion rate. We will focus on the long-wavelength modes of the cosmological perturbations, by neglecting any of their spatial gradients.

In the uniform-density gauge, the perturbation of the local expansion rate vanishes completely in the long-wavelength limit. In particular, at the first order, it is given by $\delta H = \nabla^2 f$, where $f$ is a function of the scalar metric perturbations. Physically, the fluid behaves over the surfaces of constant density as if it contains a single component, without entropy fluctuations, which is why there is no long-wavelength back-reaction effect when the local expansion rate is averaged over these surfaces.

Hence, taking $F=H$ and $\delta H \simeq 0$ in (\ref{main}), we get
\begin{align}\begin{split} \label{result}
\frac{\expval{H}_\chi}{H_0} \,&=\, 1 \,+\, \frac{\Hu'' + 3\Hu\Hu' - 5\Hu^3}{2\Hu}\,\langle(\delta\eta)^2\rangle \\[5pt]
&\hspace{20pt} +\, \frac{\Hu' - \Hu^2}{2\Hu} \dv{}{\eta} \langle(\delta\eta)^2\rangle \,.
\end{split}\end{align}
Here, we have also assumed that the average of $\delta\eta$ vanishes over the surfaces of constant density, $\expval{\delta\eta} = 0$, which implies that its average over the surfaces of constant clock field is
\be
\expval{\delta\eta}_\chi \,=\,  3\Hu\,\langle(\delta\eta)^2\rangle + \frac12 \dv{}{\eta} \langle(\delta\eta)^2\rangle \,.
\ee

Let us now consider the case where the fluid contains one component which dominates all others, and has a background equation of state parameter $w$ which is approximately constant. In that case, the background expansion rate in comoving coordinates is
\be
\Hu \, = \, \frac{\alpha}{\eta} \, = \, \frac{2}{(3w + 1) \eta} \, ,
\ee
and so (\ref{result}) yields
\begin{align}\begin{split} \label{result-dominant-fluid}
\frac{\expval{H}_\chi}{H_0} \,&=\, 1 - \frac{(5\alpha-2)(\alpha+1)}{2\eta^2}\langle(\delta\eta)^2\rangle  \\[5pt]
&\hspace{20pt} - \frac{\alpha + 1}{2\eta} \dv{}{\eta} \langle(\delta\eta)^2\rangle\,.
\end{split}\end{align}

Similarly, let's assume that the clock field has a constant equation of state parameter $w_\chi$, and an energy density which is small compared to the dominant fluid. Since the conformal time $\eta_\chi$ and the clock field are both constant over the same surfaces, it follows from (\ref{clock-field}) that
\be
\rho_\chi \, = \, \rho_{0\chi}(\eta_\chi) \, = \, \rho_{0\chi}(\eta) + \delta\rho_\chi \, ,
\ee
where $\delta\rho_\chi$ is the density fluctuation of the clock field, when its background value is expressed in terms of the background conformal time. Expanding at the first order, with $\eta_\chi = \eta - \delta\eta$, we get
\be\label{delta-eta}
\delta\eta \, = \, -\frac{\delta\rho_\chi}{\rho_{0\chi}'} \, .
\ee

Assuming that the clock field does not exchange energy with the dominant fluid, its background energy density satisfies the equation
\be \label{clock-field-background}
\rho_{0\chi}' + 3\Hu (1 + w_\chi) \rho_{0\chi}\, = \, 0\,.
\ee
In the long-wavelength limit, the density fluctuation $\delta\rho_\chi$ also satisfies the previous equation, as shown in the appendix at the end of this paper. The fractional density fluctuation in the clock field is then approximately constant. Hence we can define the constant
\be
\delta C \, = \, \frac{1}{3(1+w_\chi)}\,\frac{\delta \rho_\chi}{\rho_{0\chi}}\,,
\ee
and so the expression (\ref{delta-eta}) for $\delta\eta$ becomes
\be
\delta\eta \, = \, \frac{\delta C}{\alpha}\,\eta\,.
\ee
We can now evaluate both terms in (\ref{result-dominant-fluid}), and obtain
\be
\frac{\expval{H}_\chi}{H_0} \,=\, 1 - \frac{15}{4}(1+w)\,\langle(\delta C)^2\rangle \,.
\ee

Hence, we see that the leading back-reaction effect does lead to a decrease of the locally-measured expansion rate as long as
\be
w \, > \, -1 .
\ee
Since for the scalar field matter sources considered in previous works $-1 < w < 1$, our result agrees with those in \cite{Vacca, Leila}, namely that the back-reaction of cosmological perturbations leads to a decrease in the locally-measured expansion rate.

\section{Conclusions}

We have studied the back-reaction of super-Hubble density fluctuation modes on the average value of the locally measured expansion rate, $H$, working to leading order in perturbation theory.  In contrast to previous studies, we consider matter to be a perfect fluid rather than a scalar field.  We consider the average of the locally measured expansion rate, averaging over surfaces of space-time which have a constant value of a clock field, a field whose effect on the evolution of space-time is negligible (in the same way that today the radiation has a negligible effect on the expansion of space). Working to leading order in the gradient expansion, we have shown that, provided that the clock field has fluctuations relative to surfaces of constant density,  then the super-Hubble modes lead to a decrease in the average value of $H$ compared to what it would be in the absence of fluctuations. 

The back-reaction effect of super-Hubble modes thus has the same effect as a negative contribution to the cosmological constant would. This is most easily seen in the formalism of \cite{ABM} where the back-reaction effect is described by an effective energy-momentum tensor. Since on super-Hubble scales spatial gradient terms are negligible, and (according to linear cosmological perturbation theory) the perturbation modes are frozen in and hence no temporal gradient terms appear,  we know that this effective energy-momentum tensor must take on the form of a cosmological constant which is time-dependent if the phase space of super-Hubble modes changes\footnote{The non-conservation of the effective energy-momentum tensor of fluctuations is due to an energy exchange between background and fluctuations. As shown in \cite{Abramo}, the full Bianchi identities are satisfied to the leading order in which we are calculating.}. Applied to an accelerating background where the phase space of infrared modes is increasing since modes are exiting the Hubble horizon, we find that the magnitude of the back-reaction effect is increasing in time, indicating a relaxation of the initial cosmological constant which is causing the accelerated expansion \cite{RHBbrRev}.

There is a clear physical reason why the average expansion rate decreases in the presence of fluctuations: a matter density perturbation induces a gravitational potential well.  For super-Hubble modes, the negativity of the gravitational energy exceeds the positivity of the matter energy, leading to a net decrease in the effective energy density. Note that for sub-Hubble modes the sign of the back-reaction effect is expected to be opposite. However, in an accelerating background with $w \simeq -1$, the phase space of the sub-Hubble modes is constant (and hence can be renormalized to zero), whereas that of super-Hubble modes is a secularly increasing effect.

Our analysis has been done to leading order in the cosmological perturbations, namely it represents an effect which is quadratic in the amplitude of the fluctuations.  It would be interesting to study the effect at higher orders, to see if the secularly growing negative contribution to the locally measured expansion rate persists (the evidence from the work of \cite{Leila} supports this conjecture). If it were so, the back-reaction effect could then lead to a dynamical relaxation of the cosmological constant as conjectured in \cite{RHBbrRev}.

\section*{Acknowledgments}

VC is supported in part by FRQNT. The research at McGill is supported in part by funds from NSERC and from the Canada Research Chair program.  RB is grateful for the hospitality of Lavinia Heisenberg at the Institute for Theoretical Physics of the ETH Z\"{u}rich, and support by a Pauli Fellowship.

\section*{Appendix to section III}

In the third section of this paper, we assume that the space-time contains two perfect fluids, a dominant one and a clock field $\chi$, whose energy densities are denoted $\rho_w$ and $\rho_\chi$ respectively. These fluids also have constant equations of state parameters $w$ and $w_\chi$. The total energy density and pressure for the two fluids are thus
\begin{align}\begin{split}
\rho \, &= \, \rho_w + \rho_\chi \,, \\[5pt]
p \, &= \, w\,\rho_w + w_\chi\,\rho_\chi \,.
\end{split}\end{align}
In the uniform-density gauge used in this paper, $\delta\rho = 0$, and so $\delta\rho_w = -\delta\rho_\chi$. The total pressure fluctuation is then 
\be
\delta p \, = \, (w_\chi - w)\,\delta\rho_\chi\, .
\ee

Furthermore, we assume that the two fluids do not exchange energy with one another. Conservation of energy for the clock field then implies that
\be
u_\chi^\A \,\nabla_\A \rho_\chi + \nabla_\A u_\chi^\A \,(1+w_\chi)\,\rho_\chi \, = \, 0 \,,
\ee
where $u_\chi^\A$ is the normalized four-velocity of the clock field, which might differ from the four-velocity of the dominant fluid. At leading order in the perturbations,
\be
u_\chi^0 \,=\, \frac{1}{a}\left(1 + \frac12\phi\right) \,,
\ee
where $\phi$ is the time-time component of the metric fluctuations. In the long-wavelength limit, and in the uniform-density gauge, the linearized Einstein equations imply that
\be
\frac12 \phi \,=\, \frac{\delta p}{\rho + p} \,=\, \frac{w_\chi - w}{(1 + w_\chi) \rho_\chi + (1+w)\rho_w}\,\delta\rho_\chi\,.
\ee

In that limit and gauge, it can also be shown \cite{Comeau} that the expansion rate of the clock field coincides with the background expansion rate, $\nabla_\A u_\chi^\A = 3H_0$. The conservation of energy equation thus becomes, at the first order in the perturbations,
\be
\delta\rho_\chi' + 3\Hu\left(1 + w_\chi + \frac{w - w_\chi}{ 1 + \frac{1+w}{1+w_\chi}\,\frac{\rho_w}{\rho_\chi}}\right)\delta\rho_\chi \,=\, 0\, .
\ee
When the dominant fluid dominates the energy density in the background, $\rho_w \gg \rho_\chi$, the previous equation simplifies to
\be
\delta\rho_\chi' + 3\Hu\left(1 + w_\chi \right)\delta\rho_\chi \,=\, 0\, .
\ee
Hence, in this case, the perturbation $\delta\rho_\chi$ does obey the same equation as the background energy density of the clock field, namely (\ref{clock-field-background}), as stated in the third section of this paper.

\end{document}